\documentclass[preprints,article,accept,moreauthors,pdftex]{mdpi} 

\firstpage{1} 
\pubvolume{1}
\issuenum{1}
\articlenumber{0}
\pubyear{2021}
\copyrightyear{2020}
\datereceived{} 
\dateaccepted{} 
\datepublished{} 
\hreflink{https://doi.org/} 
\pdfoutput=1

\Title{Velocity multistability vs ergodicity breaking \\in a biased periodic potential}

\TitleCitation{Velocity multistability vs ergodicity breaking in a biased periodic potential}

\Author{Jakub Spiechowicz $^{1}$*\orcidA{}, Peter H{\"a}nggi$^{2}$\orcidC{} and Jerzy {\L}uczka $^{1}$\orcidB{}}

\AuthorNames{Jakub Spiechowicz, Peter H{\"a}nggi and Jerzy {\L}uczka}

\AuthorCitation{Spiechowicz J.; H{\"a}nggi P. and {\L}uczka J.;}

\address[2]{%
$^{1}$ \quad Institute of Physics, University of Silesia in Katowice, 41-500 Chorz{\'o}w, Poland,\\ 
$^{2}$ \quad Institute of Physics, University of Augsburg, Universitätstr. 1,
86135 Augsburg, Germany}

\corres{Correspondence: jakub.spiechowicz@us.edu.pl}

\abstract{Multistability, i.e. the coexistence of several attractors for a given set of  system parameters is one of the most important phenomena occurring in dynamical systems. We consider it in velocity dynamics of a Brownian particle driven by thermal fluctuations and moving in a biased periodic potential. In doing so we focus on the impact of ergodicity - a concept which lies at the core of statistical mechanics. The latter implies that a single trajectory of the system is representative for the whole ensemble and as a consequence the initial conditions of the dynamics are fully forgotten. The ergodicity of the deterministic counterpart is \emph{strongly} broken and   we discuss how the velocity multistability depends on the starting position and velocity of the particle. While for non-zero temperature the ergodicity is in principle restored, in the low temperature regime the velocity dynamics is still affected by initial conditions due to \emph{weak} ergodicity breaking. For moderate and high temperature the multistability is robust with respect to the choice of the starting position and velocity of the particle.}


\begin{document}
\maketitle

\section{Introduction}

Research in nonequilibrium statistical physics provides a wealth of intriguing dynamics in which phenomena  that are forbidden in equilibrium states  may emerge. Prominent examples include anomalous diffusion \cite{metzler2014, spiechowicz2017scirep, spiechowicz2019chaos, spiechowicz2019njp}, Brownian yet non-Gaussian diffusion \cite{leptos2009, wang2012, metzler2017, spiechowicz2020pre2, barkai2020}, noise-assisted transport \cite{hanggi2009,slapik2019prappl} and negative mobility \cite{machura2007,nagel2008,slapik2019,slapik2020}, to name only a few. While the behaviour of low dimensional systems, where usually only one or two attractors rule the dynamics, has been studied intensively, much less is known for systems where several attractors coexist for a given set of the system parameters. This feature called multistability is found commonly in different areas of science such as physics, chemistry, biology, economy and in nature \cite{pisarchik2014}. 


In this paper we reinvestigate in this context the paradigmatic model of nonequilibrium statistical physics, namely, underdamped Brownian motion in a biased periodic potential. This nonlinear system enjoys never ending interest as its different aspects have been studied already for several decades \cite{risken, lindner2001, reimann2001a, reimann2001b, constantini1999, lindenberg2005, marchenko2014, kramer2013,lindner2016, zhang2017,marchenko2017, cheng2018, goychuk2019, spiechowicz2020pre, goychuk2021, spiechowicz2021pre2}. The latter are mostly focused on the diffusive properties of the system. For instance, it may exhibit unusual phenomena like the giant diffusion \cite{reimann2001a,reimann2001b,lindner2016,spiechowicz2021pre2} or the non-monotonic temperature dependence of a diffusion coefficient    \cite{lindner2016,marchenko2019,spiechowicz2020pre,spiechowicz2021pre2}. Both these effects are related to a bistability observed in the velocity dynamics of the system. The later effect is well known since the work by Risken et al. \cite{vollmer1983} who found that at low friction and appropriate bias values the velocity can be stable in a locked solution (the particle is trapped in a potential minimum) but also in a running solution (the motion is unbounded in space).

Here we focus on  multistability of the Brownian velocity dynamics in a tilted periodic potential. Despite so many years of intensive research on various aspects of this setup the latter peculiar effect has been addressed only very recently \cite{spiechowicz2020pre} and later it was explained by recoursing to the arcsine law \cite{spiechowicz2021pre}, which is a cornerstone of extreme-value statistics. Specifically, we investigate the role of ergodicity breaking and its consequences on the velocity multistability. Ergodicity lies at the basis of statistical mechanics and implies that over long enough observation times, the time averages of observables correspond to the equilibrium ensemble averages \cite{metzler2014,meroz2015,bouchaud1992}. Equivalently, it states that a single trajectory is representative for the ensemble. An increasing number of systems exhibit nonergodic properties \cite{metzler2014,meroz2015,bouchaud1992}, in particular due to the ultra slow dynamics and non-exponential relaxation.

The paper is organized as follows. In Sec. 2 we recall the formulation of the model and introduce the dimensionless quantities. In the next section we discuss the results, in particular the effect of ergodicity breaking on the velocity multistability occurring in this paradigmatic system. Finally, Sec. 4 provides a discussion and concluding remarks.

\section{Methods}

In this work we study dynamics of a classical inertial Brownian particle of mass $M$ moving in a spatially periodic and symmetric potential $U(x) = U(x + L)$ of  period $L$ and subjected to a static bias $F$. This system can be described by the following Langevin equation
\begin{equation}
	\label{model}
	M\ddot{x} + \Gamma\dot{x} = -U'(x) + F + \sqrt{2\Gamma k_B T}\,\xi(t), 
\end{equation}
where the dot and prime denote differentiation with respect to the time $t$ and the particle coordinate $x$, respectively. The parameter $\Gamma$ is the friction coefficient, $T$ is temperature and $k_B$ denotes the Boltzmann constant. We consider the potential $U(x)$ in the form
\begin{equation}
	\label{potential}
	U(x) = -\Delta U \sin{\left(\frac{2\pi}{L}x\right)}, 
\end{equation}
where $\Delta U$ denotes \textit{half} of the potential barrier height. Thermal equilibrium fluctuations are modeled by the $\delta$-correlated Gaussian white noise whose statistical characteristics read
\begin{equation}
	\langle \xi(t) \rangle = 0, \quad \langle \xi(t)\xi(s) \rangle = \delta(t-s).
\end{equation}
The noise prefactor $2 \Gamma k_B T$ satisfies the fluctuation-dissipation theorem that ensures the canonical Gibbs statistics when the system is at the equilibrium state.

The above Langevin equation (1) can be transformed into the dimensionless form
\begin{equation}
	\label{dimless-model}
	\ddot{\hat x} + \gamma \dot{\hat x} = -\mathcal{U}'(\hat x)+ \sqrt{2\gamma \theta}\,\hat{\xi}(\hat t)
\end{equation}
by introducing the rescaled coordinate $\hat{x}$ and time $\hat{t}$, 
\begin{equation}
	\label{scaling}
	\hat x = \frac{2\pi}{L} x, \quad \hat t = \frac{t}{\tau_0}, \quad 
	\tau_0 = \frac{L}{2\pi} \sqrt{\frac{M}{\Delta U}}, 
\end{equation}
where the characteristic time $\tau_0$ is proportional to 
 the inverse of frequency $\omega_0$ of small oscillations in the potential well of $U(x)$. 
The effective dimensionless potential is 
\begin{equation}
\mathcal{U}({\hat x}) = -\sin {\hat x} -f {\hat x}. 
\end{equation}
The dimensionless friction coefficient $\gamma$ and bias $f$ read
\begin{equation}
	\gamma = \frac{1}{2\pi}\frac{L}{\sqrt{M \Delta U}}\, \Gamma,  \quad f = \frac{1}{2\pi}\frac{L}{\Delta U} F. 
\end{equation}
The rescaled temperature $\theta$ is the ratio of thermal energy $k_B T$ to half of the barrier height the particle needs to overcome the original potential well, namely,
\begin{equation}
	\theta = \frac{k_B T}{\Delta U}.
\end{equation}
The dimensionless thermal noise $\hat{\xi}({\hat{t}})$ is statistically equivalent to $\xi(t)$ meaning that it is a stationary Gaussian stochastic process with vanishing mean. Later we use only the rescaled quantities and therefore in order to improve the readability of notation from now on we omit the hat appearing in Eq. (\ref{dimless-model}).

The model of a Brownian particle moving in a washboard potential formulated in terms of the Langevin equation (\ref{dimless-model}) for decades served as a tool for investigation of transport effects occurring in both classical and quantum systems. For instance, it has been employed for understanding the dynamics of phase across the Josephson junction \cite{junction}, rotating dipoles in external fields \cite{coffey}, superionic conductors \cite{gruner1981}, charge density waves \cite{fulde1975} as well as cold atoms dwelling in optical lattices \cite{denisov2014,kindermann2017,dechant2019}. Yet other systems are mentioned in Ref. \cite{risken}.

The analytical methods of solution for the Fokker-Planck equation corresponding to Eq. (\ref{dimless-model}) are not yet elaborated, therefore in doing so, we rely solely on precise numerical simulations. All calculations have been done using a Compute Unified Device Architecture (CUDA) environment implemented on a modern desktop graphics processing unit (GPU). This method allowed to speed up necessary calculations by a factor of the order $10^3$ as compared to the traditional methods. We refer interested reader to Ref. \cite{spiechowicz2015cpc,seibert2011} where more details on this scheme can be found.

\section{Results}

In this paper we investigate various aspects of multistability in the velocity dynamics of a Brownian particle dwelling in a tilted periodic potential. This interesting phenomenon has been addressed for this setup only very recently \cite{spiechowicz2021pre}, however, it has been reported also in systems driven by other types of noise. Examples include fractional Gaussian noise \cite{goychuk2019}, Ornstein-Uhlenbeck and harmonic Levy noise \cite{li2020,li2021}, to name but a few.
\begin{figure}[t]
	\centering
	\includegraphics[width=0.66\linewidth]{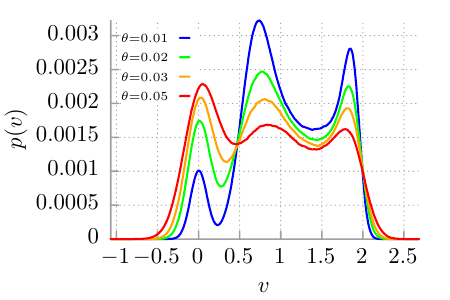}
	\caption{The probability distribution $p(v)$ for the instantaneous long time velocity $v=v(t)$ of the Brownian particle is illustrated for $t=10^4$ and selected values of  temperatures $\theta$ of the system. The used parameters read $\gamma = 0.66$ and $f = 0.91$.}
	\label{fig1}
\end{figure}

In Fig. \ref{fig1} we exemplify the velocity multistability phenomenon occurring in this system. The probability distribution $p(v)$ for the instantaneous long time velocity $v$  is depicted there for different dimensionless temperatures $\theta$. The issue of measurement of the instantaneous velocity of a Brownian particle is presented in Ref. \cite{raizen}. 
In Fig. \ref{fig1}, one can observe  three well pronounced maxima.  One of them corresponds to the velocity $v = 0$ (the locked state) and the other two with $v \neq 0$ are related to running solutions. It means that these values occur significantly more frequently than the others and therefore are more stable. This observation matches the common definition of multistability for stochastic systems \cite{grebogi}. We note that as temperature $\theta$ increases the difference between each maximum becomes less pronounced and eventually disappears.
\begin{figure}[t]
	\centering
	\includegraphics[width=0.66\linewidth]{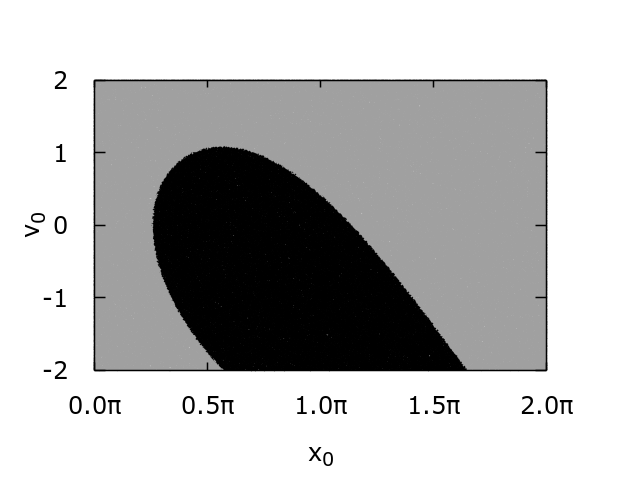}
	\caption{The basins of attraction for the time averaged velocity $\mathbf{v}$ of the particle. The black color codes the locked state $\mathbf{v} = 0$ whereas the grey part indicates the regime with running solutions $\mathbf{v} \neq 0$. Parameters read $\gamma = 0.66$, $f = 0.91$ and $\theta = 0$.}
	\label{fig2}
\end{figure}
\begin{figure}
	\centering
	\includegraphics[width=0.48\linewidth]{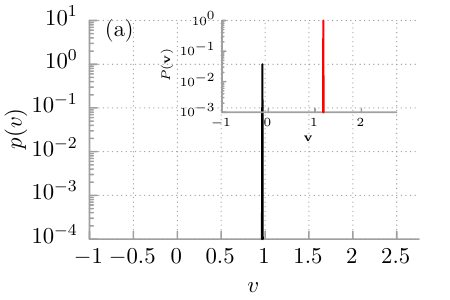}
	\includegraphics[width=0.48\linewidth]{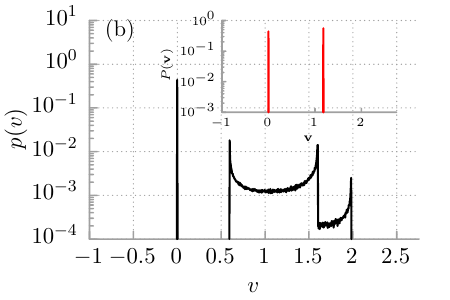}\\
	\includegraphics[width=0.48\linewidth]{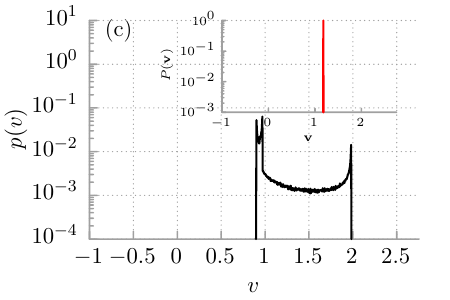}
	\includegraphics[width=0.48\linewidth]{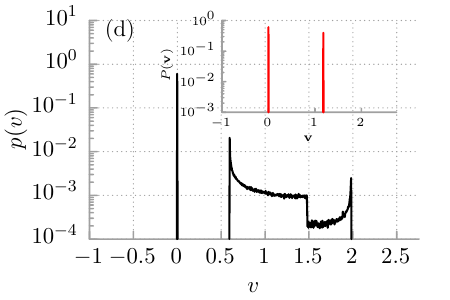}\\
	\includegraphics[width=0.48\linewidth]{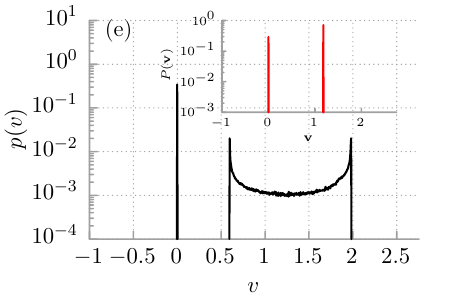}
	\includegraphics[width=0.48\linewidth]{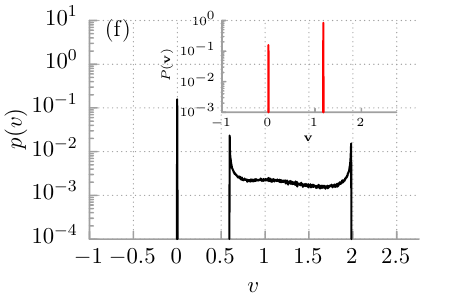}
	\caption{The probability distribution $p(v)$ for the instantaneous long time velocity $v$ of the Brownian particle is depicted in the deterministic regime $\theta = 0$ for $t = 10^4$ and different choice of the initial conditions for the system. Panel (a): $p_{x_0}(x) = \delta(x)$, $p_{v_0}(v) = \delta(v)$; (b): $p_{x_0}(x) = \mathsf{U}(0,2\pi)$, $p_{v_0}(v) = \delta(v)$; (c): $p_{x_0}(x) = \delta(x)$, $p_{v_0}(v) = \mathsf{U}(-2,2)$; (d): $p_{x_0}(x) = \delta(x - \pi)$, $p_{v_0}(v) = \mathsf{U}(-2,2)$; (e): $p_{x_0}(x) = \mathsf{U}(0,2\pi)$, $p_{v_0}(v) = \mathsf{U}(-2,2)$; (f): $p_{x_0}(x) = \mathsf{N}(0,1)$, $p_{v_0}(v) = \mathsf{N}(0,1)$, where $\mathsf{U}(a,b)$ indicates the uniform distribution over the interval $[a,b]$. Likewise $\mathsf{N}(\mu,\sigma^2)$ is the Gaussian distribution with the mean $\mu$ and the variance $\sigma^2$. In the inset the corresponding probability distribution $P(\mathbf{v})$ for the time averaged velocity $\mathbf{v}$ is shown. Parameters read $\gamma = 0.66$, $f = 0.91$, $\theta = 0$.}
	\label{fig3}
\end{figure}

In Ref. \cite{spiechowicz2021pre} the origin of the multistability effect  is explained in terms of the arcsine law for the velocity dynamics at the zero temperature limit $\theta = 0$, i.e. as the trace of deterministic dynamics perturbed by thermal noise.  In such a case in the long time regime the velocity $v(t)$ of the particle is a time-periodic function. Moreover, the ergodicity of the setup is strongly broken which means that its phase space can be divided into two non-intersecting invariant sets corresponding to the locked and running state \cite{spiechowicz2016scirep}. We visualize this in  Fig. \ref{fig2},  where  the time averaged particle velocity 
\begin{equation}
	\mathbf{v} = \lim_{t \to \infty} \frac{1}{t} \int_0^t ds \, \dot{x}(s)
\end{equation}
is depicted as a function of the initial conditions for the coordinate $x(0) = x_0$ and velocity $v(0) = v_0$. The black region corresponds to the locked state with $\mathbf{v} = 0$ whereas the grey one indicates the regime of a running solution for which $\mathbf{v} \neq 0$. Therefore different initial conditions $\{x_0,v_0\}$ can lead to a distinct average velocity $\mathbf{v}$. It is a disturbing situation as typically in experiments the initial conditions are not known \emph{a priori} or can be settled only with a finite resolution. To get rid of the dependence of the obtained results on the initial conditions one needs to average over them. In Ref. \cite{spiechowicz2021pre} the authors distributed $x_0$ and $v_0$ uniformly over the intervals $[0,2\pi]$ and $[-2,2]$, respectively. Moreover, they found that in such a case the initial conditions induce an almost uniformly distributed phase shift $\varphi$ in the time-periodic dependence of the velocity $v(t)$ in the long time regime. This in turn results in the arcsine law for the velocity probability density $p(v)$ which constitutes the backbone of multistability in this system.

In this work we present a complementary study. Namely, we investigate in detail the influence of various distributions of initial conditions $\{x_0,v_0\}$ on the velocity multistability phenomenon.  In Fig. \ref{fig3} we show the probability distribution $p(v)$ for the instantaneous long time velocity $v$ of the  Brownian particle for the deterministic system $\theta = 0$ and different choice of the initial conditions. In simulations the moment of time is fixed $t_i = 10^4$. In the inset we depict the corresponding probability distribution $P(\mathbf{v})$ for the time averaged velocity $\mathbf{v}$. In panel (a) the initial position and velocity are fixed, $x_0=0, v_0=0$.    The corresponding probability densities are represented by the Dirac-delta $p_{x_0}(x) = \delta(x)$ and $p_{v_0}(v) = \delta(v)$, respectively. Consequently, since the system is noiseless $\theta = 0$ the resulting probability distributions $p(v)$ and $P(\mathbf{v})$ for the instantaneous long time $v$ and time averaged velocity $\mathbf{v}$, respectively, take  the   Dirac-delta forms.  All phase space trajectories follow the same route and the multistability effect is absent. The situation changes drastically already if the initial position of the particle is distributed uniformly over the period $L = 2\pi$ of the potential $U(x)$, i.e. $p_{x_0}(x) = \mathsf{U}(0,2\pi)$, see panel (b). Here $\mathsf{U}(a,b)$ indicates the uniform distribution over the interval $[a,b]$. The starting velocity of the particle can be fixed $p_{v_0}(v) = \delta(v)$ but still the systems displays multimodality in the probability density $p(v)$. In fact, in such a case even four distinct maxima are visible there. In the inset we note that both  locked $\mathbf{v} = 0$ and  running $\mathbf{v} = 0$ states are represented in the ensemble of system trajectories. If we permute the initial conditions, i.e. the starting coordinate $p_{x_0}(x) = \delta(x)$ but $p_{v_0}(v) = \mathsf{U}(-2,2)$, see panel (c), the multistability emerges but the locked state is not sampled at all. This situation can be modified depending on the choice of the initial coordinate as it is demonstrated in panel (d) where in contrast $p_{x_0}(x) = \delta(x - \pi)$.

Overall, if the ergodicity of the system is broken the initial conditions are never forgotten and therefore crucially impact the results. Depending on the circumstances this behaviour may be seen as \emph{a feature, not a bug}. Nevertheless, the only way to cure it is to \emph{properly} average over the initial conditions. In doing so each of them must be taken into account equally and none can be preferred. This requirement translates to the fact that initial conditions must be equally probable and therefore uniformly distributed over the whole phase space. Since the system under consideration is spatially periodic $U(x) = U(x + L)$, the periodic boundary condition can be employed to yield $p_{x_0}(x) = \mathsf{U}(0,2\pi)$. It is not the case for the starting velocity $v_0$ of the particle which in principle is unbounded. However, naturally such a situation cannot be implemented in numerical simulations and therefore one needs to carefully check the impact of the initial velocity subspace volume on the obtained results. As we demonstrated, if this is not done thoroughly one can significantly spoil the outcomes and e.g. break the inherent symmetries of the system \cite{kostur2008}. We checked that in the considered regime the condition $p_{v_0}(v) = \mathsf{U}(-2,2)$ is sufficient and further increase of the initially chosen velocity subspace volume does not alter the outcomes. In Fig. \ref{fig3} (e) we reproduce the result from Ref. \cite{spiechowicz2021pre} obtained for $p_{x_0}(x) = \mathsf{U}(0,2\pi)$ and $p_{v_0}(v) = \mathsf{U}(-2,2)$. The characteristic U-shape part which portrays the arcsine law corresponding to the running state is visible in the probability density $p(v)$. Consequently, the velocity dynamics is multistable. 

One can claim that the initial conditions, especially the velocity, should be distributed according to the Gaussian probability density since then it obeys the canonical Gibbs statistics (Maxwell-Boltzmann distribution) valid for equilibrium systems. Obviously such a choice does not satisfy the above discussed condition of equal probability. In panel (f) we show that as a consequence of the non-uniformity for $p_{x_0}(x) = \mathsf{N}(0,1)$ and $p_{v_0}(v) = \mathsf{N}(0,1)$, where $\mathsf{N}(\mu,\sigma^2)$ is the Gaussian distribution with the mean $\mu$ and the variance $\sigma^2$, the results are deformed and the arcsine law is not properly recovered. There is one more argument that the condition of equal probability is the only one 
correct and consistent with the case of non-zero temperature. In the running state the long time velocity trajectory $v(t)$ is a periodic function of time and can be well approximated by the simple periodic function \cite{spiechowicz2021pre} 
\begin{equation}
	\label{approx}
	V(t) = A\sin{(\omega t + \phi)} + c. 
\end{equation}
For a fixed set of the system parameters, the constants $(A, \omega, c)$ are the same for all initial conditions $\{x_0, v_0\}$. However, the distribution of the phase shift $\phi$ depends on the distribution of initial conditions $\{x_0, v_0\}$. This fact is reflected in different probability densities $p(v)$ for the instantaneous velocity depicted in Fig. \ref{fig3}. Since the ergodicity of the system is broken the distributions $p(v)$ generally depend on the measurement time $t = t_i$. The exception is the uniform distribution for the phase $\phi$ corresponding to the panel (e) in Fig. \ref{fig3} for which $p(v)$ is time-invariant \cite{spiechowicz2021pre}. The latter feature is characteristic for ergodic systems and is crucial from the experimental point of view. 
\begin{figure}[t]
	\centering
	\includegraphics[width=0.49\linewidth]{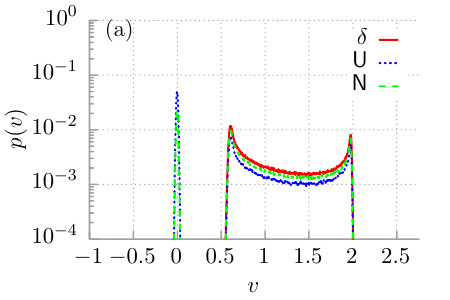}
	\includegraphics[width=0.49\linewidth]{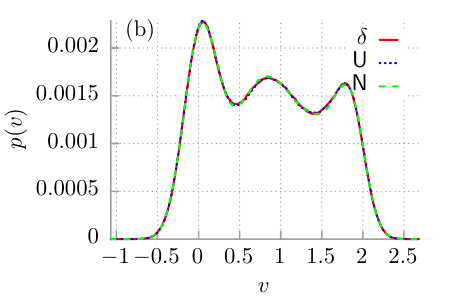}
	\caption{The probability distribution $p(v)$ for the instantaneous long time velocity $v$ of the Brownian particle is depicted for $t = 10^4$ and different initial conditions of the system. The red solid line indicates $p_{x_0}(x) = \delta(x)$, $p_{v_0}(v) = \delta(v)$. The blue dotted line corresponds to $p_{x_0}(x) = \mathsf{U}(0,2\pi)$, $p_{v_0}(v) = \mathsf{U}(-2,2)$. The green dashed line denotes $p_{x_0}(x) = \mathsf{N}(0,1)$, $p_{v_0}(v) = \mathsf{N}(0,1)$. In panel (a) temperature is $\theta = 0.0001$ while in (b) $\theta = 0.05$. Other parameters read $\gamma = 0.66$, $f = 0.91$.}
	\label{fig4}
\end{figure}

As we just reported the ergodicity of the deterministic system with $\theta = 0$ is broken. One may argue that the case $\theta = 0$ is only an idealization and in practise there exists no realistic situation with zero temperature. 
However, the ergodicity breaking in a deterministic system often carries prominent consequences also for non-zero temperature. In particular, for any positive temperature $\theta > 0$ the system described by Eq. (\ref{dimless-model}) is ergodic, although it is not a trivial fact since it is driven by  noise \cite{cheng2015}. At low temperature the whole phase space is accessible due to  thermally activated escape events connecting the coexisting deterministic disjoint attractors. Nevertheless the time $\tau$ after it is fully sampled becomes extremely long. If the temperature tends to zero $\theta \to 0$ it goes to infinity $\tau \to \infty$. Such a situation is often termed as weak ergodicity breaking \cite{bouchaud1992,spiechowicz2016scirep}. In the latter case the initial conditions do not fade but in fact modify the results. We exemplify this feature in Fig. \ref{fig4} (a) where we depict the probability distribution $p(v)$ for the instantaneous long time velocity $v$ of the Brownian particle for different initial conditions and low temperature $\theta = 0.0001$. Clearly, when the particle starts from $x_0 = 0$ and $v_0 = 0$, see the red solid line, even in the long time limit there are only running solutions. On the other hand, if the initial position $x_0$ and velocity $v_0$ are either uniformly or normal distributed, see the blue or green line, respectively, the multistability emerges but still one can note quantitative difference between these two initial conditions. In contrast, in panel (b) we depict the same characteristics but for higher temperature $\theta = 0.05$. Then thermal fluctuations are strong enough to recover the ergodicity of the system and no longer there are differences between different initial conditions. Even when the particle trajectories starts from the same point in the phase space $x_0 = 0$, $v_0 = 0$, see the red solid line, the whole density is obtained.

\section{Discussion}

In conclusion, we thoroughly investigated the influence of initial conditions distribution on the multistability of velocity dynamics for the Brownian particle in a tilted periodic potential. The ergodicity of the deterministic system is strongly broken and therefore the initial conditions are never forgotten and crucially impact the obtained results. The only way to \emph{correctly} sample the whole state space of the system is to \emph{average} over them in such a way that no single one is preferred. The latter condition translates to an \emph{uniform} distribution in the initial phase space of the system. We demonstrated that while for non-zero temperature the ergodicity is in principle restored, in low temperature regime the results are still significantly affected by the initial conditions due to the weak ergodicity breaking. It means that the time needed for the ergodicity reinstatement tends to infinity when temperature goes down to zero. For moderate and high temperature regime the detected multistability is robust with respect to the choice of initial conditions. It is valid even when the whole ensemble starts from a given point in the phase space of the system. A remaining question is how the time needed for the ergodicity restoration depends on temperature. This constitutes, however, a challengeable objective which we hope to address in the future. 






\vspace{6pt} 



\authorcontributions{J.S performed all the calculations. All authors contributed to the analysis and discussion of the results as well as writing up of this work.}

\dataavailability{The data that support the findings of this study are available from the corresponding author upon reasonable request.}

\acknowledgments{This work has been supported by the Grant NCN No. 2017/26/D/ST2/00543 (J.S.). Moreover, P.H. and J.S. wish to acknowledge the financial support obtained via the University of Augsburg Guestprogramme funded by Bavarian Ministery for Science and Art, Germany.}

\conflictsofinterest{The authors declare no conflict of interest.}

\end{paracol}
\reftitle{References}

\end{document}